\documentclass [a4paper]{article}

\usepackage{amsmath}
\usepackage[latin1]{inputenc}
\usepackage{graphicx}
\usepackage{psfrag}
\usepackage{float}

\pdfpageattr{/Group <</S /Transparency /I true /CS /DeviceRGB>>} 
\usepackage[svgnames]{xcolor}
\usepackage{tikz}
\usepackage{pgfplots}
\PassOptionsToPackage{cmyk}{xcolor}
\usepackage{amsmath}
\usepackage{caption}
\usepackage{subcaption}
\usepackage{units}
\usepackage{marvosym}
\usepackage{textcomp}
\usetikzlibrary{calc,fadings,decorations}
\usetikzlibrary{decorations.pathmorphing}
\usetikzlibrary{decorations.pathreplacing}
\usetikzlibrary{shapes.symbols}
\usetikzlibrary{shadows} 
\usetikzlibrary{spy}
\usetikzlibrary{patterns}
\usetikzlibrary{circuits}
\usetikzlibrary{intersections}
\usetikzlibrary{positioning}
\usetikzlibrary{plotmarks}
\usetikzlibrary{circuits.ee.IEC}
\definecolor{lenscolor}{rgb}{0.8,1,1}
\definecolor{probe}{rgb}{0,0,1}
\usepgfplotslibrary{groupplots}
\usepgfplotslibrary{external}
\tikzset{external/force remake}
\pgfkeys{/pgf/number format/.cd,
        1000 sep={}}
\pgfplotsset{clip marker paths=true,compat=newest}
\usepackage{lmodern}

\begin{document}

\title{The resting potential of nerve cells and the Na,K-pump}

\author{Josef J\"ackle\thanks{E-mail: josef.jaeckle@uni-konstanz.de}\\
Department of Physics\\
University of Konstanz\\
78457 Konstanz, Germany}

\maketitle

\begin{abstract}
In this pedagogical paper a coherent explanation of the resting potential of nerve cells is given in terms of its determining factors. These are the currents of active transport of the ions to which the membrane is permeable, their membrane permeabilities and their concentrations in the extracellular fluid. They play the role of the independent variables in the problem and simultaneously also determine the concentrations of the permeating ions inside the cell. The resting state is assumed to be a stationary state, and the resting potential is understood to be the membrane potential in this state. The explanation presented here differs from the conventional one found in textbooks with respect to the handling of the intracellular concentrations of the permeant ions. In the conventional description the causal relations between these and the independent variables are not taken into account, and the intracellular concentrations are treated like additional independent variables. The "causal" theory presented here 
leads to a simple formula for the resting potential in terms of the currents of active transport (pumping by the Na,K-pump and carrier transport) and the membrane permeabilities for sodium and potassium ions. In addition, it leads to two results not obtained in the conventional treatment: An upper bound exists for the magnitude of the resting potential, and the difference between the resting potential and the diffusion potential, which can be measured after poisoning of the ion pumps, is small.

\end{abstract}

\section{Introduction}

According to the conventional view (Tuckwell 1988; L\"auger 1991; Somjen 2004) the resting potential consists of two parts. The first part arises from the differences between the extra- and intracellular concentrations of the permeant ions in the absence of active transport. This potential can be measured after the Na,K-pumps have been poisoned with an agent like ouabain. After the blocking of the pumps the membrane potential relaxes quickly to a value determined by the extra- and intracellular concentrations. It then remains constant for a longer time, during which the intracellular concentrations practically are still those of the resting state. This potential can be identified as the diffusion potential ($V_{diff}$) of the resting state without pumping. The Goldman-Hodgkin-Katz formula  (G-H-Kformula) is usually accepted as a valid expression. This formula is a weighted average over the Nernst potentials of the three types of permeant ions, to which K$^+$ makes the largest contribution because it has the 
highest membrane permeability. 

A second part of the resting potential ($\Delta V_m$) only occurs if the net current of active transport, which is the sum of the currents of active transport for Na$^+$ and K$^+$, is different from zero. While $V_{diff}$ depends on the currents of active transport only implicitly via the intracellular concentrations, $\Delta V_m$ depends on these also explicitly. It is often assumed that the net current of active transport is not very different from the net current produced by the Na,K-pumps alone. In this case the condition for a non-zero second part $\Delta V_m$ is that the net pump current does not vanish, i.e. that the pumps are "electrogenic" (L\"auger 1991). This condition is fulfilled for the Na,K-pump. $\Delta V_m$ is in good approximation linearly proportional to the net pump-current density. Note that here the pumping of Na$^+$ and that of K$^+$ have the same weight. The second part $\Delta V_m$ can be measured as the change of the membrane potential from the resting potential $V_m$ to the 
diffusion potential $V_{diff}$, which occurs after poisoning of the ion pumps:
\begin{equation}
\Delta V_m = V_m - V_{diff}.
\end{equation}
It turns out that $\Delta V_m$ is usually much smaller in magnitude than $V_{diff}$. Therefore, the diffusion part, which can be evaluated more easily than the difference $\Delta V_m$, often is a sufficient approximation also to the resting potential. However, in the conventional treatment of the resting potential, I cannot find an argument that would explain this relative smallnes of $\Delta V_m$. But such an argument can be given in the treatment presented here. It is a consequence of the large asymmetry between the membrane permeabilities for Na$^+$ and K$^+$.

The paper is organized as follows. In Sec.2  from a particular process of preparation of the cell's resting state essential conclusions concerning the role of the Na,K-pump in the generation of the resting potential are drawn. They are formulated in three conjectures in Section 3. The first two of them can be proven independently of a specific model for passive ion transport (Section 4). The third conjecture can be proven mathematically only on the basis of a specific model for passive ion transport, which quantitatively defines the property of {\it permeability} for the different ions. Here the third conjecture is proven using Goldman's electrodiffusion model for passive ion transport (Goldman 1943). This crude approximation leads to a simple and plausible result for $V_m$ in terms of active-transport currents and permeabilities (Section 5). Section 6 is concerned with the difference between the resting potential and the diffusion potential. 
The difference is shown to be small, because active transport of K$^+$-ions has only a small influence on the resting potential due to the relatively high potassium permeability. In Sec.7 it is shown that an upper bound to the magnitude of the resting potential exists. This result is similar to Hoppensteadt and Peskin's finding for an alternative Ohmic-conduction model. Their model is reviewed briefly in an Appendix.
\footnote{The present paper has a predecessor in (J\"ackle 2007). There a calculation similar to that of Sec.5 is presented, which, however, differs in two important respects. In the present version the mathematics is more transparent thanks to the consistent limitation to Na$^+$-, K$^+$- and Cl$^-$-ions as the permeant ions both inside and outside the cell, and it is not assumed that the ion currents due to active transport are unique functions of the extra- and intracellular concentrations and the membrane potential. Instead, the actual values of these currents are assumed to be known.}

\section{Preparation of the resting state}   

The need of a negative membrane potential for a stationary resting state of a nerve cell follows from the conditions of stable electro-mechanical equilibrium as applied to a cell into which a given collection of anions from the cell's metabolism is locked. The conditions are electric charge neutrality inside the cell and osmotic-pressure balance at the cell walls. A particular process of generating the resting state is considered, which fulfils these conditions. The process consists of the following sequence of steps. At the start, let the cell be filled with the extracellular fluid, and the concentrations of Na$^+$, K$^+$- and Cl$^-$-ions be the same inside and outside the cell. After the final cell volume $v_C$ has been decided on, a given number $N_i$ of ions of variable negative charge, to which the cell membrane is impermeable, is implanted in the cell. The mean valence $z_i$ of these ions is negative and has a value close to -1. Their concentration $c_i$ is given by $N_i/v_C$. At the next step $|z_i|\
cdot N_i$ chloride ions are removed from the cell to neutralize the negative charge $z_i\cdot e N_i$ of the implanted ions. ($e$ is the elementary electric charge.) This step restores electric charge neutrality in the cell, but disturbs osmotic pressure balance since the total number of ions is changed. An exception is the special case $z_i= -1$, when the valences of the ions exchanged are equal on average. In general, this step increases the total number of ions in the cell by $(1-|z_i|) N_i$ if $|z_i|<1$, and decreases it by $(|z_i|-1)N_i$ if $|z_i|>1$. Keeping the cell electrically neutral, osmotic pressure balance can be restored by removing $(1-|z_i|)N_i/2$ neutral pairs of a cation and a chloride ion if $|z_i|<1$, and by adding $(|z_i|-1)N_i/2$ such pairs if $|z_i|>1$. Altogether, $(1+|z_i|)N_i/2$ chloride ions have been removed in both cases of $|z_i|$, while $(1-|z_i|)N_i/2$ cations have been removed from the cell in the case $|z_i|<1$, but $(|z_i|-1)N_i/2$ cations have been added to the cell in the 
case $|z_i|>1$.

The cell membrane is permeable to Cl$^-$-ions. Therefore, to stabilize the resulting concentration gradient of the chloride ions, a potential step from a negative value of the potential inside the cell to a positive value outside must be built up across the membrane, to push Cl$^-$-ions out of the cell. Such a potential would not be needed if a Cl$^{-}$-pump of suitable strength could pump Cl$^{-}$-ions out of the cell. 
For ion pumping by the Na,K-pump, a potential step would also not be needed in the hypothetical case of a mean valency $z_i=+1$ of the impermeant ions, since then the cell could be kept electrically neutral by the pumping of monovalent cations out of the cell. In the case under consideration, the required membrane potential, which is negative according to our definition, not only pushes Cl$^{-}$-ions out of the cell, but also pulls positive Na$^+$- and K$^+$-ions into it. This second effect of the membrane potential must be counteracted by the {\it pumping of positive ions out of the cell}.\footnote{Apart from stabilizing the resting state, the generation of ion gradients across the cell membrane by ion pumping also provides the fuel for secondary active ion transport, in which the downhill motion of a cation is coupled to the uphill motion of another ion or an uncharged molecule.}
 Without ion pumping osmotic pressure balance could not be maintained since the electric forces implied by the potential step alone would attract too high a concentration of cations in the cell. The cell and surrounding fluid would end up only in a {\it Donnan equilibrium}  (see, e.g., Adam, L\"auger, Stark, 2009), in which the osmotic pressure in the cell is higher than outside. Without rigid cell walls to resist the mechanical stress, the cell would swell infinitely. When the activity of the ion pumps is sufficient to keep the cell volume at the chosen value, ion pumping rates and cell mobilities determine which portion of the cations in the cell are Na$^+$-, and which are K$^+$-ions.

In short: The expulsion of the Cl$^-$-ions from the cell by the implantation of the impermeant anions leads to a concentration gradient of Cl$^-$, which needs to be stabilized by a negative membrane potential. This, in turn, would attract positive ions into the cell, and therefore must be counteracted by pumping of positive ions out of the cell.

By the {\it Na,K-pump} Na$^+$-ions are pumped out of the cell, while K$^+$-ions are pumped into it. The strength of the pumping of either species is measured by the respective pump-current densities $\Phi_{Na}^{(P)}$ and $\Phi_{K}^{(P)}$ ( in mole/(m$^2$s ). Since currents out of the cell are counted as positive and currents into the cell as negative, $\Phi_{Na}^{(P)}$ is positive and $\Phi_{K}^{(P)}$ negative. Only the pumping of positive ions out of the cell stabilizes the resting state, whereas pumping into the cell has a destabilizing effect. A first guess would be that the effect of the ion pumps on the emerging membrane potential depends on the sum $\Phi_{Na}^{(P)}+\Phi_{K}^{(P)}$ of the two pump-current densities. Accordingly, only for {\it electrogenic ion pumps} (L\"auger, 1991), for which by definition this sum is non-zero, a membrane potential would exist. In this case, for the accepted pumping ratio $\Phi_{K}^{(P)}/ \Phi_{Na}^{(P)}$ = -2/3, where the minus sign expresses the fact that Na$^+$- and 
K$^+$-ions are pumped into opposite directions, the K$^+$-pumping would annul two-thirds of the effect of the Na$^+$-pumping. However, the effect of pumping also depends on the permeability of the membrane for passive backflow of the ion. Of course, under stationary conditions, for each type of ion separately, passive backflow compensates the flow generated by the ion pumps.

 .

The effect of permeability is analogous to the effect which the leakiness of a water trough has on the depth of water pumped into the trough (leakiness as permeability). It is obvious that under stationary conditions the water level in the trough is lower if the trough is more leaky. For the same rate of pumping water into the trough, a higher leakiness of the trough must be compensated by a lower water level, which goes with a lower hydrostatic pressure at the bottom of the trough. In the cell, a higher membrane permeability is compensated by a smaller gradient of the electrochemical ion potential, which goes with a smaller concentration gradient. In the case considered, the membrane at rest is much more permeable to K$^+$- than to Na$^+$-ions. Therefore the intracellular concentration of K$^+$ is enhanced by the inward pumping of these ions much less than the intracellular concentration of Na$^+$-ions is reduced by the their outward pumping. Consequently, the pumping of Na$^+$-ions should have a stronger 
effect than the pumping of K$^+$ions, and the pump-current density $\Phi_{Na}^{(P)}$ should have a larger weight than  $\Phi_K^{(P)}$ in the result for $V_m$.

\section{Three conjectures}

This analysis of the preparation process suggests the following hypotheses:
(1) A stable resting state with a negative resting potential exists if and only if ion pumping has the effect of reducing the intracellular cation concentration
relative to the extracellular one.
(2) The absolute magnitude of the resting potential is the larger the larger the effect of pumping is.
(3) The absolute magnitude of the resting potential increases not only with increasing Na$^+$pump rate, but also with decreasing Na$^+$ permeability of the membrane.
    For K$^+$, the dependence of the resting potential on these parameters is a decrease rather than an increase.

\section{Ion pumping versus electrostatic forces}

In the following it is shown that the resting potential is directly related to the reduction of the intracellular cation concentration owed to the activity of the Na,K-pump. The relation derives from the conditions of intracellular charge neutrality and osmotic-pressure balance, as discussed already in Sec.2. The only simplifying assumption made is that only Na$^+$- and K$^+$-ions participate in active transport, but Cl$^-$-ions are only subject to the electrostatic membrane potential. The fact which justifies this simplification is that the intracellular  Cl$^-$-concentration is not far from the equilibrium value, or, in other words, the Nernst potential for Cl$^-$ is not far from the resting potential.(See, e.g., Aickin, Betz and Harris, 1988.)
The relation obtained reads 
\begin{equation}\label{neu1}
(\Delta c_{+})_P / c_{+}^{(o)} = s(V_m),
\end{equation}
where $c_{+} = c_{Na} + c_K$ is the joint concentration of the cations Na$^+$ and K$^+$, $\Delta c_{+}=c_{+}^{(i)}-c_{+}^{(o)}$ 
is the change of their intracellular concentration relative to their extracellular one, and $(\Delta c_{+})_P$ is the part of this change which is caused by the activity of the ion pumps (see below). A negative value of $(\Delta c_{+})_P$  means that ion pumping leads to a reduction of the cation concentration in the cell. $c_{+}^{(o)}$ is the extracellular cation concentration.
The dimensionless function of the resting potential on the r.h.s. of eq.(\ref{neu1}), which also depends on the mean valence $z_i$, is given by 
\begin{equation}\label{g1}
s(V_m) = - exp(-FV_m/RT) - \frac{2 z_i}{1-z_i} + \frac{1+z_i}{1-z_i} exp(FV_m/RT),
\end{equation}
This function can also be written as
\begin{equation}\label{g2}
s(V_m) = \frac{2}{1-z_i} \left(\sinh(FV_m/RT) + z_i (\cosh(FV_m/RT)-1)\right ).
\end{equation}

The meaning of the relation (\ref{neu1}) is that, in order to obtain a certain value $V_m$ of the resting potential, the relative change 
$(\Delta c_{+})_P / c_{+}^{(o)}$ of the intracellular cation concentration brought about by ion pumping alone must amount to a value given by $s(V_m)$. At this point, nothing is known yet about the dependence of this concentration change on pump current densities, membrane permeabilities and other parameters, except that it is  zero in the absence of pumping. From the second form (\ref{g2}) of $s(V_m)$ it follows immediately that $s(V_m)$ for $z_i<0$ is a monotonically increasing function of $V_m$ for $V_m<0$, and goes to zero for $V_m$ going to zero. With this result the conjectures (1) and (2) are proven for the present model.

Fig. 1 shows $s(V_m)$ in the domain $V_m<0$ for three different values of $z_i$. The difference between the curves shown is small. The different curves do not intersect. As is shown below (see eq.(\ref{seq}), the small difference between the upper or lower curve and the curve in the  middle is the relative {\it total} change of the intracellular cation concentration, including the effect of the membrane potential, for the respective valencies. For the upper curve ($z_i$=-2.0) the relative change is positive, for the lower curve ($z_i$=-0.5) it is negative. (Compare with Sec.2.) Figure 1 shows clearly that the effects of ion pumping and electrostatic attraction indeed very nearly compensate one another.

\begin{figure}[!htbp]
	\centering
	\scalebox{0.8} {
	\begin{tikzpicture}[thick]
	\begin{axis}[axis lines=middle, samples=400,width=6cm,height=6cm,xmin=-65,xmax=15, domain=-100:0, xtick={25, 0, -25, -50, -75, -100},ytick={0,-5, -10, -15, -20, -25}, every tick/.style={
        black,
        ymin=-15,
        ymax=5,
        thick,
      },minor x tick num=4, minor y tick num=1,xticklabel style={yshift=0.5ex,
       anchor=south},xlabel={$V_m$ [mV]},ylabel={$s(V_m)$},ymin=-25,ymax=15,every axis x label/.style={at={(current axis.right of origin)},anchor=west}, every axis y label/.style={at={(current axis.above origin)},anchor=south}]

	\addplot [thick] {-exp(-x/25.3)-2*(-0.5)/(1-(-0.5))+(1+(-0.5))/(1-(-0.5))*exp(x/25.3)};

	\addplot [thick,dashed] {-exp(-x/25.3)-2*(-1)/(1-(-1))+(1+(-1))/(1-(-1))*exp(x/25.3)};
	
	\addplot [thick] {-exp(-x/25.3)-2*(-2)/(1-(-2))+(1+(-2))/(1-(-2))*exp(x/25.3)};
	
	\end{axis}
	\end{tikzpicture}
	}
  \caption{The function $s(V_m)$ (eq. \ref{g1}) for $V_m \leq 0$ and three different values of $z_i$ (from below: $z_i$=-0.5, -1, -2). (Temperature 20 \textdegree C)}
  \label{fig:1}
\end{figure}
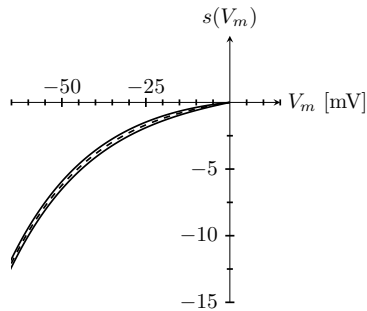

The calculation leading to the result of eqs.(\ref{neu1},\ref{g1}) goes as follows.
Let $c_l^{(o)}$ and $c_l^{(i)}$ denote the extra- and intracellular concentrations of the permeant ions Na$^+$, K$^+$ and Cl$^-$, and $c_i$ and $z_i$ the concentration and mean valence of the impermeant ions locked in the cell. The condition of electrical charge neutrality inside the cell \footnote{ Strictly speaking, electrical charge neutrality holds only for the ion concentrations in the inner ({\it bulk}) region of a cell. Near the cell wall thin layers contain the electric charges from which the membrane potential arises.}
can be written as 
\begin{equation}\label{neutr.i}
e(c_{Na}^{(i)} + c_{K}^{(i)} - c_{Cl}^{(i)})+z_i e c_i =0.
\end{equation}
The same condition is imposed on the given extracellular concentrations. This reads
\begin{equation}\label{neutr.o}
c_{Na}^{(o)} + c_{K}^{(o)} - c_{Cl}^{(o)} = 0.
\end{equation}
This equation may not be well fulfilled for the values of extracellular concentrations determined experimentally. In the blood of the giant axon of the squid, e.g., the charge of free Mg$^{++}$-ions, not taken into account here, compensates nearly 20 percent of the negative Cl$^-$ charge. However, we put up with this inaccuracy in order to keep the mathematics as transparent as possible.
Similarly, the condition of osmotic pressure balance is written as
\begin{equation}\label{osm}
R T\cdot(c_{Na}^{(i)} + c_{K}^{(i)} + c_{Cl}^{(i)} + c_i) =
R T\cdot(c_{Na}^{(o)} + c_{K}^{(o)} + c_{Cl}^{(o)}),
\end{equation}
where $R$ is the gas constant and $T$ the absolute temperature. It is advantageous to express the conditions (\ref{neutr.i}) and (\ref{osm}) in terms of the shifts of concentration inside the cell relative to the concentrations outside. They are defined as 
\begin{equation}\label{Deltac}
\Delta c_l = c_l^{(i)}-c_l^{(o)} \quad\mbox{for} \quad l = \mbox{Na, K, and Cl}.
\end{equation}
In terms of the concentration shifts $\Delta c_l$ the condition of internal charge neutrality becomes (after subtraction of eq.(\ref{neutr.o}) from
eq.(\ref{neutr.i}))
\begin{equation}\label{neutr1}
\Delta c_{Na}+\Delta c_K-\Delta c_{Cl}+z_i c_i=0,
\end{equation}
and the condition of osmotic-pressure balance eq.(\ref{osm}) reads
\begin{equation}\label{osm1}
\Delta c_{Na}+\Delta c_K+\Delta c_{Cl}+c_i=0.
\end{equation}
Multiplying eq.(\ref{osm1}) by $z_i$ and subtracting from eq.(\ref{neutr1}) eliminates $c_i$ and leads to the equation
\begin{equation}\label{zwischen1}
(1-z_i) (\Delta c_{Na}+\Delta c_K)-(1+z_i)\Delta c_{Cl}=0,
\end{equation} 
This equation can be obtained
directly from the considerations in Sec.2.
Note that for $z_i=-1$ the cation concentration inside the cell is the same as outside.

The decisive step in the derivation is the splitting of the concentration shifts $\Delta c_l $ for Na$^+$ and K$^+$ in two parts, which correspond to the two effects discussed in Sec.2: The effect of the electrostatic forces implied by the membrane potential and the effect of ion pumping.
The first term, which is denoted by $(\Delta c_l)_V$, is the concentration shift due to the membrane potential in thermal equilibrium. Since $V_m$ is defined as the potential inside the cell relative to that outside, the ratio of intra- and extracellular concentrations in equilibrium is given by the corresponding Boltzmann factor, viz.
\begin{equation}
\left(\frac{c_l^{(i)}}{c_l^{(o)}} \right)_V=\exp(-z_l FV_m/RT).
\end{equation}
This leads to
\begin{equation}\label{G2}
(\Delta c_l)_V = c_l^{(o)}(\exp(\mp F V_m/RT)-1)
\end{equation}
for Na$^+$ and K$^+$ ($z_l$=1) and Cl$^-$($z_l$=-1), respectively.
The second term is the remaining part of $\Delta c_l$, which is non-zero only if ion pumps are active at rest or have been active during excitation. Denoting it by $(\Delta c_l)_P$, the decomposition of the concentration shifts $\Delta c_l$ can be written as
\begin{equation}\label{G1}
\Delta c_l =   (\Delta c_l)_V + (\Delta c_l)_P.
\end{equation}
Putting eqs.(\ref{G1} with \ref{G2} and \ref{neutr.o}) into eq.(\ref{zwischen1}) and neglecting the second term $(\Delta c_l)_P$ for Cl$^-$ leads to the desired result.

It is interesting to notice that the difference between a curve $s(V_m)$ for an arbitrary value of $z_i$ and the curve for $z_i=-1$ amounts to the relative {\it total} change  $\Delta c_{+}$  for this value of $z_i$ of the cation concentration inside the cell compared with outside. This can be proved as follows. $z_i=-1$ is the special case where $\Delta c_{+}=0$. Therefore
\begin{equation}
s(V_m;z_i=-1)) = ((\Delta c_{+})_P/c_{+}^{(o)})_{z_i=-1} = -(\Delta c_{+})_V/c_{+}^{(o)}.
\end{equation}
Since $(\Delta c_{+})_V$ (eq.(\ref{G2})) does not depend on $z_i$, for general $z_i$
\begin{equation}\label{seq}
s(V_m;z_i)-s(V_m;z_i=-1) = (\Delta c_{+})_P/c_{+}^{(o)} + (\Delta c_{+})_V/c_{+}^{(o)} = \Delta c_{+}/c_{+}^{(o)}.
\end{equation}

In addition to the derivation of eq.(\ref{g1}), two further consequences of the conditions (\ref{neutr1} and (\ref{osm1}) of electric charge neutrality and osmotic pressure balance shall be mentioned here. The first concerns the relation between $c_i$ and $\Delta c_{Cl}$. It is obtained by subtracting eq.(\ref{neutr1}) from eq.(\ref {osm1}) and reads 
\begin{equation}\label{zwischen2}
\Delta c_{Cl}=-\frac{1-z_i}{2} c_i.
\end{equation}
This equation also follows directly from the considerations in Sec.2. Since the concentration $c_i$ must be a positive number, we obtain from eq.(\ref{zwischen2}) for valencies $z_i$ different from $+1$ the inequalities
\begin{equation}\label{ineq}
\Delta c_{Cl} < 0  \quad \mbox{for}\quad z_i < +1, \quad\mbox{and}\quad
\Delta c_{Cl} > 0  \quad \mbox{for}\quad z_i > +1, 
\end {equation}
and for $z_i=+1$ the equality
\begin{equation}\label{eq}
\Delta c_{Cl}=0.
\end{equation}
With the assumption of thermal equilibrium of the Cl$^-$-ions (i.e.: $\Delta c_{Cl}=(\Delta c_{Cl})_V$, see eq.(\ref{G2})) the same relations hold for $V_m$ alone, viz. 
\begin{equation}\label{ineq1}
V_m < 0\quad \mbox{for}\quad z_i < +1,\quad V_m > 0\quad \mbox{for}\quad z_i > +1,\quad \mbox{and}
\quad V_m=0\quad \mbox{for}\quad z_i=0. 
\end{equation}
Under the same assumption, eq.(\ref{zwischen2}) also implies that the cell volume, inversely proportional to $c_i$, is determined by $V_m$. For the special case $z_i$=-1 it gives 
\begin{equation}
c_i=-\Delta c_{Cl},
\end{equation}
which means that the impermeant ions implanted in the cell replace an equal number of Cl$^-$-ions. 


The second remark concerns the value of the mean valence $z_i$ of the impermeant ions in the cell. Solving eq.(\ref{neutr1}) for $z_i c_i$ and eq.(\ref{osm1})
for $c_i$ and dividing, the following expression for $z_i$ is obtained:
\begin{equation}
z_i=\frac{\Delta c_{+}-\Delta c_{Cl}}{\Delta c_{+}+\Delta c_{Cl}}=-\frac{1-\frac{\Delta c_{+}}{\Delta c_{Cl}}}{1+\frac{\Delta c_{+}}{\Delta c_{Cl}}}.
\end{equation}
For the ion concentrations in the giant axon of the squid (Gainer et al. (1984)) this yields $z_i = -1.0$; with a correction which takes extracellular Mg$^{++}$-ions into account one obtains $z_i=-0.8$. Typical values of $z_i$ range between -0.5 and -2.0.

\section{Pumping efficiency and membrane permeability}

The question to be answered is how the cation concentration shift $(\Delta c_+)_P$ depends quantitatively on the pump rates and on the permeabilities of the membrane for passive ion flow. It was argued already by Woodbury (1965) and by Katz (1966) that the permeability asymmetry between sodium and potassium is important for the generation of the resting potential. Qualitatively, the role played by the membrane permeabilities has been explained above in terms of a particular preparation of the resting state, and is expressed in hypothesis (3) above. How can this result be made quantitative by a mathematical treatment?

The mathematical problem is the following: There are five quantities to be determined from five equations. These quantities are
the resting potential $V_m$, the concentration $c_i$ of the impermeant ions inside the cell and the intracellular concentrations of the permeant ions Na$^+$, K$^+$ and Cl$^-$ relative to their extracellular ones ($\Delta c_{Na}$, $\Delta c_K$, $\Delta c_{Cl}$). The extracellular concentrations 
$c_{Na}^{(o)}, c_{K}^{(o)}$ and $c_{Cl}^{(o)}$ must be given. In addition, the current densities (flux densities) of active transport of these ions across the cell membrane
and the permeabilities of the membrane for passive ion transport must be known. Active transport occurs by ion pumping and secondary active transport (see, e.g. Stein 1990). Secondary active transport in most cases is transport of an ion or uncharged molecule up its concentration gradient coupled with the transport of a Na$^+$- or K$^+$-ion down their concentration gradients built up by ion pumping. Examples of such cotransport are the transfer of molecules of glucose or ions of aminoacids into the cell and the expulsion of Ca$^{++}$-ions out of the cell by coupling to the backflow of Na$^+$-ions into the cell. The current density of Na$^+$ and K$^+$ in this type of transport is denoted by $\Phi_l^{(co)}$. Since Na$^+$- and K$^+$-ions move against their direction of pumping,  $\Phi_l^{(co)}$ and $\Phi_l^{(P)}$ are of opposite sign. Their sum is smaller in magnitude than $\Phi_l^{(P)}$ alone, but still has the same sign.

As in Sec.4, the small deviation of the concentration gradient of Cl$^-$ from equilibrium with the membrane potential is neglected. With the equilibrium formula eq.(\ref{G2}) for Cl$^-$, viz.
\begin{equation}
\Delta c_{Cl} = (\Delta c_{Cl})_V = c_{Cl}^{(o)}(\exp(F V_m/RT)-1),
\end{equation}
the number of unknown quantities is reduced by one to four ($V_m, c_i, \Delta c_{Na}, \Delta c_K$). They have to be determined from four equations, which are

(i) eq.(\ref{neutr1}) for the condition of electric charge neutrality,

(ii) eq.(\ref{osm1}) for the condition of osmotic pressure balance, 

(iii) equations expressing the condition that the net currents of the Na$^+$- and K$^+$-ions across the membrane, which are the sum of the currents due to passive transport, ion pumping and secondary active transport, are zero. These equations are 
\begin{equation}\label{Phi}
\Phi_l + \Phi_l^{(P)} + \Phi_l^{(co)} = 0 \quad\mbox{for} \quad l = \mbox{Na and K},
\end{equation}
Here $\Phi_l$ and $\Phi_l^{(x)}$ denote the concentration currents of ion type $l$ in passive transport and the other modes of ion transport. Currents in outward direction are counted positive. With the abbreviation 
\begin{equation}
\Phi_l^{(a)} = \Phi_l^{(P)} + \Phi_l^{(co)}
\end{equation}
for the net current densities of Na$^+$ and K$^+$ of active transport, the equations (\ref{Phi}) can be written as
\begin{equation}\label{Phil}
\Phi_l = - \Phi_l^{(a)}.
\end{equation}
The equations (\ref{neutr1}) and (\ref{osm1}) are equivalent to the equations (\ref{zwischen1}) and (\ref{zwischen2}).

It is assumed that, for a given type of cell, the current densities $\Phi_{Na}$ and $\Phi_K$ of passive transport occurring in eqs.(5.22) are unique functions of the extra- and intracellular concentrations of Na$^+$ and K$^+$, and of $V_m$. These functions define the constitutive equations for passive transport of the permeant ions. It is further assumed that, when these functions are inserted in eqs.(5.22), these equations can be solved for 
$\Delta c_{Na}$ and $\Delta c_K$ at given $\Phi^{(a)}_{Na}$ and $\Phi^{(a)}_K$ (Mackey 1975). Unfortunately, these functions differ widely for different types of cell, and are not of simple form. This reflects the great diversity of ion channels in cell membranes (Hille 2001; Roux, 1999; Cooper et al., 1985).
However, one may ask if it is necessary to take all details of the constitutive equations into account if one is interested only in a particular property of a stationary resting state. One may think of replacing, for each type of ion, the voltage-dependent permeabilities of different kinds of ion channels by an 
{\bf effective permeability} of the whole membrane in the resting state. Following this idea, we have recourse to the constitutive equation for $\Phi_l$ proposed by Goldman (1943), to define such an effective membrane permeability, leaving open the problem of how to calculate it from detailed channel data. 
An alternative is the Ohmic-conduction model due to Hoppensteadt and Peskin (1992), which leads to an effectively similar result. It is briefly reviewed in an appendix.

Goldman's expression is an 
approximation for a model of electrodiffusion in a continuous medium under constant electric field. It is of simple form and, for ion type l with valence $z_l$, reads
\begin{equation}
\Phi_l= P_l \frac{z_l F V_m}{RT}
\frac{c_l^{(i)} exp(z_l F V_m/RT)-c_l^{(o)}}{\exp(z_l F V_m/RT)-1}.
\end{equation}
The constant $P_l$ is the membrane permeability for ion type $l$.\footnote{The definition of a permeability which is used here is due to Goldman (1943) and Hodgkin and Katz (1949) (see also, e.g., Adam, L\"auger, Stark, 2009). It has the dimension of a velocity and is measured in units of m/s. In (J\"ackle, 2007) permeabilities are defined differently with the dimension of a mobility, i.e. velocity/force, to be measured in units of s/kg. The two definitions are related by 
\begin{equation}
(P_l)_{J\ddot{a}ckle 2007} = \frac{d}{k_{B}T} P_l,
\end{equation}
where $d$ is the membrane thickness, $k_{B}$ is Boltzmann's constant, and $T$ is the absolute temperature. In the G-H-K potential formula the difference of definitions does not matter.} Solving eqs.(5.22) for the intracellular concentrations $c_{Na}^{(i)}$ and $c_K^{(i)}$ at given current densities $\Phi_l^{(a)}$ for active transport now yields for the concentration shifts
\begin{equation}
\Delta c_l = c_l^{(o)} (\exp(-F V_m/RT)-1) - (1-\exp(-F V_m/RT))\frac{RT}{F V_m}\frac{\Phi_l^{(a)}}{P_l},
\end{equation}
which is of the form of eq.(\ref{G1}). The pumping term $(\Delta c_l)_P$
can be identified with
\begin{equation}\label{DeltaP}
(\Delta c_l)_P = -(1-\exp(-F V_m/RT)) \frac{RT}{F V_m}\frac{\Phi_l^{(a)}}{P_l}.
\end{equation}
It depends linearly on the net current densities $\Phi_l^{(a)}$ of active ion transport and the inverse of the permeabilities $P_l$. This is a simple and 
plausible result. It should be noted that expression (\ref{DeltaP}) also depends on $V_m$ and temperature $T$.

Inserting these expressions into eq.(\ref{zwischen1}) with eq.(\ref{neutr.o}) yields, after division by $(RT/F V_m)\cdot (1-\exp(-F V_m/RT))$, the final result 
\begin{equation}\label{newf}
V_m \cdot f(V_m) = -\frac{RT}{F} \Psi
\end{equation}
with 
\begin{equation}\label{f}
f(V_m) = 1 + \frac{1+z_i}{1-z_i}\exp(F V_m/RT)
\end{equation}
and 
\begin{equation}\label{Psi}
\Psi = \frac{1}{c_+^{(o)}}\left(\frac{\Phi_{Na}^{(a)}}{P_{Na}} + \frac{\Phi_K^{(a)}}{P_K}\right).
\end{equation}
The dimensionless quantity $\Psi$ contains a weighted average over the current densities $\Phi_{Na}^{(a)}$ and $\Phi_K^{(a)}$ of active transport of sodium and potassium, with weighting factors proportional to the inverse of the membrane permeabilities $P_{Na}$ and $P_K$. If the Na,K-pump is regarded as the cause of the resting potential, these weighting factors may be interpreted as expressions of the {\it efficiency of the pumping} of either ion type in the generation of the resting potential. Following this interpretation, as a combination of pumping strengths and pumping efficiencies, the quantity $\Psi$ may be adequately called the {\bf "effective-pumping factor".}

In the special case $z_i=-1$ the function $f(V_m)$ is unity so that $V_m$ is strictly proprtional to $\Psi$. For arbitrary negative valence $z_i$ it can be shown that the function $V_m f(V_m)$ occurring on the l.h.s. of eq.(\ref{newf}) is a monotonically increasing function of $V_m$ in the range $V_m<0$ of interest.\footnote{This follows from the fact that $|(1+z_i)/(1-z_i)| < 1$ for $z_i<0$.} For $V_m$ going to zero this function also goes to zero. Therefore, the inverse function, which expresses $V_m$ as a function of $\Psi$, is also a monotonically increasing function of its argument, and goes to zero with it. This confirms the hypothesis (3) of Sec.3, but {\bf two qualifications are necessary}. First, the pump current densities $\Phi_l^{(P)}$ have to be replaced by the net current densities $\Phi_l^{(a)}$ of the Na,K-pumps, which include the backflow of sodium and potassium ions via processes of secondary active transport (cotransport). The net current densities $\Phi_l^{(a)}$ are smaller in magnitude 
than the bare pump-current densities $\Phi_l^{(P)}$ because in cotransport the cations in general move in the direction opposite to the pumping direction. The degree of reduction can be appreciable and can fluctuate with time, depending on the 
metabolic and electric activity of the cell. For a rough estimate, however, it may be sufficient to take only pumping into account. Secondly, a limit is set to the increase of the magnitude $|V_m|$ of the resting potential with increasing $\Phi_{Na}^{(a)}$ by the condition that the intracellular Na$^+$-concentration must not vanish. This condition is considered separately in Section 7.

In Fig.2a the function 
$V_m\cdot f(V_m)$ is plotted for two values of $z_i$ different from -1.
It is seen that the correction factor $f(V_m)$ has little effect, when $|V_m|$ is in the usual range
of about $50mV$ and larger. Fig.2b indicates a graphical method of solving eq.(\ref{newf}) for given r.h.s.

Apparently, the solution \footnote{Eq.(\ref{newf}) has a second, spurious solution for $V_m$ in the ranges $z_i<-1$ and $z_i>+1$, which are not in agreement with the inequalities (\ref{ineq1}).} of eq.(\ref{newf}) for $V_m$ depends only weakly on the precise value of $z_i$ in the usual range of $z_i$ around -1. However, over the full range of possible $z_i$-values, including the hypothetical case of positive values, $V_m$ shows greater variation. As is expressed by the inequalities eq.(\ref{ineq1}), $V_m$ vanishes when $z_i$ is raised to the valency +1 of the cations Na$^+$ and K$^+$, and becomes positive at still higher values of $z_i$. The result of a vanishing resting potential in the (hypothetical) case $z_i=+1$ is in accord with the considerations in Sec.2.

With eqs.(5.32-34)for $V_m$, eqs.(4.16) and (5.20) for $c_i$, eqs.(5.26) for $\Delta c_{Na}$ and $\Delta c_K$ the solution ($V_m, c_i, \Delta c_{Na}, \Delta c_K$) 
of the problem is found. Apart from the extracellular concentrations $c_{Na}^{(o)}, c_K^{(o)}$ and $c_{Cl}^{(o)}$, it depends on the current densities $\Phi_{Na}^{(a)}$ and$\Phi_{K}^{(a)}$ and the permeabilities $P_{Na}$ and $P_K$, which unfortunately are not known so far. (It should be remembered that according to the definition used $\Phi_{Na}^{(a)}$ is positive and $\Phi_{K}^{(a)}$ is negative.)
The solution depends only weakly on the potassium current $\Phi_{K}^{(a)}$ if $P_K$ is much larger than $P_{Na}$. This is normally the case with nerve cells, and explains an important experimental observation, as described in the next Section.

The asymmetry of the permeabilities can be assessed by calculating backward from experimental data for $V_m, \Delta c_{Na}$ and $\Delta c_K$. 
From eqs.(5.27) we obtain
\begin{equation}
\frac{P_K\cdot \Phi_{Na}^{(a)}}{P_{Na}\cdot \Phi_{K}^{(a)}} = \frac{(\Delta c_{Na})_P}{(\Delta c_{K})_P}.
\end{equation}
From eqs.(\ref{G1}) and (\ref{G2}) we find
\begin{equation}
(\Delta c_K)_P = \Delta c_K - (\Delta c_K)_V = c_K^{(i)} - c_K^{(o)} \exp(-F V_m/RT),
\end{equation}
and similarly for $(\Delta c_{Na})_P$. With the experimental data for the giant axon of the squid as a representative example,
 $c_{Na}^{(o)}=$440, $c_{Na}^{(i)}=$49, $c_K^{(o)}=$22, $c_K^{(i)}=$410 (all in mmole/l), $V_m=$-65 mV, and a temperature of 20°C, a ratio
\begin{equation}
\frac{(\Delta c_{Na})_P}{(\Delta c_{K})_P}=-46
\end{equation}
is found. This large value underlines the predominant role played by the pumping of sodium. Assuming that the current ratio $\Phi_{K}^{(a)}/ \Phi_{Na}^{(a)}$ of active transport is equal to the pumping ratio $\Phi_{K}^{(P)}/ \Phi_{Na}^{(P)}$ =-2/3, one obtains for the permeability ratio of potassium and sodium 
\begin{equation}
\frac{P_K}{P_{Na}} = 31,
\end{equation}
which agrees reasonably well with a value of 25 estimated in a different way by Hodgkin and Katz (1949).

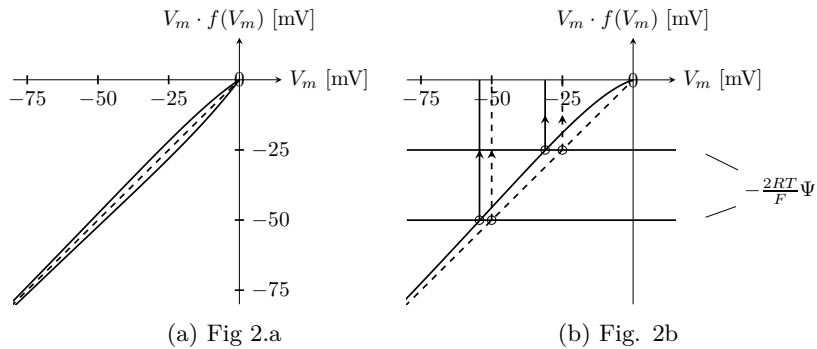
\begin{figure}[H]
\centering
\begin{subfigure}{.5\textwidth}
	\flushleft
	\scalebox{0.8} {
  \begin{tikzpicture}[thick]
\begin{axis}[axis lines=middle, samples=400,width=6cm,height=6cm,xmin=-80,xmax=15,xtick={25, 0, -25, -50, -75, -100},ytick={-25, -50, -75}, every tick/.style={
        black,
        thick,
      },domain=-100:0.1,cycle list name=color list,xlabel={$V_m$ [mV]},ylabel={$V_m \cdot f(V_m)$ [mV]},ymin=-80,ymax=15, every axis x label/.style={at={(current axis.right of origin)},anchor=west}, every axis y label/.style={at={(current axis.above origin)},anchor=south},yticklabel style={anchor=west,xshift=0.15cm}]

	\addplot+[thick, black] {x*(1+(1+(-0.5))/(1-(-0.5))*exp(x/25.3))};

	\addplot+[thick, dashed, black] {x*(1+(1+(-1))/(1-(-1))*exp(x/25.3))};
	
	\addplot+[thick, black] {x*(1+(1+(-2))/(1-(-2))*exp(x/25.3))};

	\node at (axis cs: 0,0) {$0$};
	
	\end{axis}

	\end{tikzpicture}
	}
  \caption{Fig 2.a}
  \label{fig:sub1}
\end{subfigure}%
\hspace{-1cm}
\begin{subfigure}{.5\textwidth}
	\scalebox{0.8} {
	\begin{tikzpicture}[thick]
	\begin{axis}[axis lines=middle, samples=400,width=6cm,height=6cm,xmin=-80,xmax=15, domain=-100:0, ytick=\empty, xtick={25, 0, -25, -50, -75, -100},every tick/.style={
        black,
        thick,
      },xlabel={$V_m$ [mV]},ylabel={$V_m \cdot f(V_m)$ [mV]},ymin=-80,ymax=15,every axis x label/.style={at={(current axis.right of origin)},anchor=west}, every axis y label/.style={at={(current axis.above origin)},anchor=south}]

	\addplot [thick, dashed] {x*(1+(1+(-1))/(1-(-1))*exp(x/25))};

	\addplot [thick] {x*(1+(1+(-5))/(1-(-5))*exp(x/25))};

	\draw [thick] (axis cs: 20, -25) -- (axis cs: -80, -25) node[right, pos=0] (A) {};
	\draw [thick] (axis cs: 20, -50) -- (axis cs: -80, -50) node[right, pos=0] (B) {};
	\draw (axis cs: -31.1, -25) circle (2pt);
	\draw (axis cs: -25, -25) circle (2pt);
	\draw (axis cs: -50, -50) circle (2pt);
	\draw (axis cs: -54.3, -50) circle (2pt);
	\node at (axis cs: 0,0) {$0$};
	\begin{scope}[thick,decoration={
    	markings,
	    mark=at position 0.5 with {\arrow{stealth}}}
    	] 
	    \draw[postaction={decorate}] (axis cs: -31.1, -25)  --( axis cs: -31.1,0);  
    	\draw[dashed,postaction={decorate}] (axis cs: -25, -25)  --( axis cs: -25,0);  
    
	    \draw[postaction={decorate}] (axis cs: -54.3, -50)  --( axis cs: -54.3,0);  
    	\draw[dashed,postaction={decorate}] (axis cs: -50, -50)  --( axis cs: -50,0);  
	\end{scope}
	\end{axis}

\node[below right=of A,yshift = 0.8cm,xshift=-0.5cm] (A2)  {$-\frac{2RT}{F} \Psi $};
\draw[thin] (A) -- (A2);
\draw[thin] (B) -- (A2);
	\end{tikzpicture}
	}
  \caption{Fig. 2b}
  \label{fig:sub2}
\end{subfigure}
\label{fig:test}
\end{figure}

Fig.2a:
The function $V_m \cdot f(V_m)$ occurring on the l.h.s. of eq.(\ref{newf}) is shown for $z_i=-0.5$ (lower full line) and
$z_i=-2$ (upper full line). The dashed line is for $z_i=-1$, where $f(V_m)$ is unity.

Fig.2b:
Illustration of a graphical method of solution of eq.(\ref{newf}). The dependence of $V_m$ on $z_i$ is shown by comparing the cases of $z_i=-5$ (full lines) and $z_i=-1$ (dashed lines). For two different values of the effective-pumping factor, for which $V_m$ is -25 mV and -50 mV at $z_i=-1$, the absolute value of $V_m$ at $z_i=-5$ is larger by 25 percent and 10 percent, respectively.

\section{Resting potential and diffusion potential}

It is shown here that the weak dependence of the solution ($V_m, c_i, \Delta c_{Na}, \Delta c_K$) on the rate of active transport by potassium as compared with its dependence on the same rate for sodium, which results from a high asymmetry of the permeabilities of the two cation species, viz. $P_K\gg P_{Na}$, 
explains a common experimental observation in nerve cells. It is usually found that the membrane potential changes only little when ion pumping is stopped by poisoning of the Na,K-pumps with ouabain. After poisoning, the (negative) potential increases only slightly, by about five percent, say, above its previous value.
\footnote{An exception is the giant neuron of the marine mollusc {\it Anisodoris nobilis}, for which a rise of the membrane potential by 25 percent, from -63 mV to -47 mV, was observed after ion pump blocking. (Gorman and Marmor (1970); see also Tuckwell (1988)). Possibly this exception is due to a less pronounced asymmetry between $P_K$ and $P_{Na}$.} 
The argument goes as follows. One compares the solution ($V_m, c_i, \Delta c_{Na}, \Delta c_K$) for given $\Phi_{Na}^{(a)}$ and $\Phi_{K}^{(a)}$ with a second solution
($V_m, c_i, \Delta c_{Na}, \Delta c_K$)',  which obtains for the same $\Phi_{Na}^{(a)}$, but with a different $\Phi_{K}^{(a)}$ , which makes the sum of the two vanish:
\begin{equation}
\Phi_{Na}^{(a)} +(\Phi_{K}^{(a)})' = 0.
\end{equation}
The second solution differs only little from the first, the resting potential in particular:
\begin{equation}\label{D1}
(V_m)'\approx V_m.
\end{equation}
If at first $|\Phi_{K}^{(a)}|$ is smaller than $\Phi_{Na}^{(a)}$, as is the case for the pump currents, $|\Phi_{K}^{(a)}|$ thereby increases, which makes $\Psi$ and $|V_m|$ smaller.

In the second case the net current of all active transport is zero. In a stationary state then the net current of passive transport by all
types of permeant ions is also zero. This is just the condition under which the membrane potential is given by the diffusion potential, which is determined by the extra- and intracellular concentrations of the permeant ions. It is denoted by $V_{diff}$. Usually it is expressed by the G-H-K-potential formula. Therefore in the second case the equality
\begin{equation}\label{D2}
(V_m)' = (V_{diff})'
\end{equation}
holds.

When, after starting from the original state, all active transport is switched off by poisoning, the intracellular ion concentrations for a while remain the same.   The membrane potential is then given by the diffusion potential for the concentrations of the first solution, which is denoted by $V_{diff}$.
Since the intracellular concentrations of the permeant ions differ only little, the diffusion potentials for the two cases are similar, too:
\begin{equation}\label{D3}
(V_{diff})' \approx V_{diff}.
\end{equation}

Combining the three statements (\ref{D1}), (\ref{D2}) and (\ref{D3}) one concludes that
\begin{equation}
V_m \approx V_{diff},
\end{equation}
which means that the difference between the membrane potential before and after the poisoning of the ion pumps is only small. In the argument, it was tacitly assumed that poisoning of the pumps also stops cotransport with Na$^+$- and K$^+$-ions, if this is not negligible anyway.

\section{Upper bound for magnitude of resting potental}

The result given by eqs.(\ref{newf}-\ref{Psi}) would predict an unlimited increase of the magnitude $|V_m|$ of the resting potential with increasing pump rate. However, for the model considered, the condition that the sodium concentration $c_{Na}^{(i)}$ in the cell must not vanish leads to an upper limit for the pump rate. This also sets a limit to $|V_m|$ .\footnote{It should also be borne in mind that, under the assumption of a purely passive distribution of the Cl$^-$-ions, by eq.(\ref{zwischen2}) an upper bound for $|V_m|$ implies a lower bound to the cell volume $v_C$.}

The connection between $c_{Na}^{(i)}$ and $V_m$ is as follows. From eq.(\ref{DeltaP}) we get
\begin{equation}\label{6.1}
c_{Na}^{(i)} = c_{Na}^{(o)}exp(-FV_m/RT) + (exp(-FV_m/RT)-1)\frac{RT}{FV_m}\frac{\Phi_{Na}^{(a)}}{P_{Na}}.
\end{equation}
To express the dependence of $V_m$ on the pump rate we consider $\Phi_{Na}^{(a)}$ as a measure of the pump rate, and set in eq.(\ref{Psi}) for potassium
\begin{equation}\label{zwischen2a}
\Phi_{K}^{(a)}= -r\cdot \frac{P_K}{P_{Na}}\Phi_{Na}^{(a)},
\end{equation}
with $r$ independent of the pump rate. Because of the permeability imbalance $P_{Na}\ll P_K$, the coefficient $r$ is but small ($0\le r\ll 1$). Eq.(\ref{newf}) then becomes
\begin{equation}\label{6.2}
(FV_m/RT) \cdot f(V_m) = -\frac{1}{c_{+}^{(o)}}(1-r)\frac{\Phi_{Na}^{(a)}}{P_{Na}}.
\end{equation}
Elimination of the sodium pumping factor $\Phi_{Na}^{(a)}/P_{Na}$ in eq.(\ref{6.1}) via eq.(\ref{6.2}) yields the desired expression:
\begin{equation}\label{6.3}
\frac{c_{Na}^{(i)}}{c_{Na}^{(o)}}\hspace{0.2cm}=\hspace{0.2cm}exp(-FV_m/RT)\hspace{0.2cm} -\hspace{0.2cm} \frac{c_{+}^{(o)}}{c_{Na}^{(o)}}\frac{1}{1-r}(exp(-FV_m/RT)-1)\cdot f(V_m).
\end{equation}
With the abbreviation
\begin{equation}\label{defa}
\frac{c_{+}^{(o)}}{c_{Na}^{(o)}}\frac{1}{1-r} = a >1
\end{equation}
this can be written as 
\begin{equation}\label{6.4}
\frac{c_{Na}^{(i)}}{c_{Na}^{(o)}} = -(a-1)exp(-FV_m/RT) \hspace{0.2cm} + \hspace{0.2cm}  a \left(1-\frac{1+z_i}{1-z_i}(1-exp(FV_m/RT))\right).
\end{equation}
Since $a>1$, this function of $V_m$ for negative $V_m$ goes to zero at a certain value of $V_m$, which is the smallest value of $V_m$ (i.e. the largest value of its absolute magnitude $|V_m|$) that is compatible with a non-negative $c_{Na}^{(i)}$. Its absolute value should be considered as an unattainable upper limit, since the pumping rate of the Na,K-pump goes to zero with the intracellular Na$^+$-concentration. In other words, the pumping rate cannot be pushed to the point where no Na$^+$-ions would be left in the cell. Fig.3 shows a plot of this function for three different values of the mean valence $z_i$. The constant $a$ was estimated from data for the giant axon of the squid. (a=1.073). The zeroes of the function are found to be at -59 mV, -68 mV and -75 mV for $z_i$=-0.5, -1 and -2.respectively. ( $RT/F$ =25.3 mV at a temperature of 20°C. ) This is only slightly larger in magnitude than typical resting potentials in neurons. \footnote{ The observation that in Fig.3 the curve for $z_i$=-2 has a 
maximum at an intermediate value of $V_m$ can be explained by the fact that for $z_i < -1$ the number of Cl$^-$-ions pushed out of the cell for the sake of electroneutrality exceeds the number of impermeant ions replacing the Cl$^-$-ions. To keep osmotic pressure balanced, cations are pulled back into the cell. For low pumping rate, and with a coefficient $a$ (eq.(\ref{defa})) only slightly larger than 1, this also affects the Na$^+$-ions. (See Sec.2.)}

\begin{figure}[!htbp]
    \centering
    \scalebox{0.8} {
    \begin{tikzpicture}[thick]
    \begin{axis}[axis lines=middle, samples=400,width=7cm,height=7cm,xmin=-80,xmax=5, domain=-100:0, xtick={0, -10, -20, -30, -40, -50, -60, -70, -80}, ytick={0, 0.25, 0.5, 0.75, 1}, every tick/.style={
        black,
        ymin=0,
        ymax=1.2,
        thick,
      },minor x tick num=0, minor y tick num=0,yticklabel style={anchor=west,xshift=0.25cm},xticklabel style={yshift=0.5ex,
       anchor=north},xlabel={$V_m$ [mV]},ylabel={$c_{\mathrm{Na}}^{(\mathrm{i})}/c_{\mathrm{Na}}^{(\mathrm{o})}$},ymin=-25,ymax=15,every axis x label/.style={at={(current axis.right of origin)},anchor=west}, every axis y label/.style={at={(current axis.above origin)},anchor=south}]

    \addplot [thick, dashed] {-0.073*exp(-x/25.3)+1.073*(1-1/3*(1-exp(x/25.3)))};
    
    \addplot [thick] {-0.073*exp(-x/25.3)+1.073*(1-0*(1-exp(x/25.3)))};
    
    \addplot [thick, dashed] {-0.073*exp(-x/25.3)+1.073*(1-(-1/3)*(1-exp(x/25.3)))};   
    
    \addplot[mark=+,mark size=5,thick] coordinates {
        (-65, 0.11)
    };
    
    \node (Z) at (axis cs:0,0) {};   
    
    \end{axis}
    
    \node at (Z) {$0$};       
    
    \end{tikzpicture}
    }
  \caption{The ratio $c_{\mathrm{Na}}^{(\mathrm{i})}/c_{\mathrm{Na}}^{(\mathrm{o})}$ for sodium ion depletion in the cell over resting potential $V_m$. $z_i=-0.5,-1,-2$ from right to left.}
  \label{fig:3}
\end{figure}
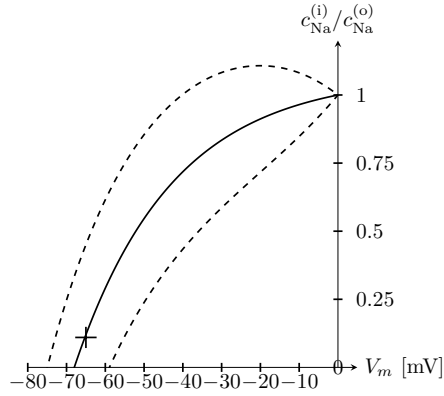

The validity of the relation eq.(\ref{6.3} or \ref{6.4}) between membrane potential $V_m$ and depletion ratio $c_{Na}^{(i)}/c_{Na}^{(o)}$ in the resting state can be tested by comparison with measured data. For the giant axon of the squid the depletion ratio is 49/440=0.11 and the resting potential is -65 mV, which agrees very well with this equation for a plausible value of the  mean valence $z_i$ of -1. The corresponding data point, marked by + in Fig.3, lies very precisely on the theoretical curve for this value of $z_i$. The perfect agreement with the theoretical curve is unexpected in view of the inaccuracy of the experimental data and the simplifications of the theory. 

It is interesting to note that the experimental value of $|V_m|$ is only 3 percent below its theoretical upper limit, while the depletion ratio for the Na$^+$-ions still amounts to 11 percent. This suggests that the absolute values of the resting potential in general tend to be close to the upper limit predicted by theory.

\section{Summary}

The paper begins with a non-mathematical analysis of a particular process of generating the resting state, from which qualitative conclusions regarding the necessity of a negative membrane potential and the neutralizing activity of cation pumps can be drawn. The process starts with the cell being filled with the extracellular solution of the permeant ions and the mixture of organic ions which are connected with the cell's metabolism. These ions are locked in the cell. They are negatively charged on average. The final state of the process has to fulfil the conditions of a stable electro-mechanical equilibrium, i.e. the conditions of electric charge neutrality inside the cell and of osmotic-pressure balance at the cell walls. The validity of the conclusions, which can be drawn from this process of preparation of the resting state, is subsequently checked by calculations for a pump-leak model, in which only the most abundant permeant ions Na$^+$, K$^+$ and Cl$^-$ are taken into account. The currents of active 
transport are the sum of the currents of ion pumping and the currents in 
modes of secondary active transport. It is assumed that active transport of Cl$^-$-ions can be neglected, so that the chloride ions are passively distributed between inside and outside the cell. 

First it is shown that a monotonic relation exists between the absolute magnitude of the resting potential and the necessary reduction of the intracellular cation concentration by active transport. This result does not depend on a specific model for passive ion transport. 

The second result concerns the dependence of the resting potential on the currents of active transport of Na$^+$ and K$^+$ and their membrane permeabilities.
The expectation of an increase of the absolute magnitude of the resting potential with increasing strength of active transport and with decreasing membrane permeability of the Na$^+$-ions, and of the opposite dependences for the K$^+$-ions, is confirmed by a calculation using the Goldman-Hodgkin-Katz expression for passive ion currents with suitably chosen effective permeabilities. The formula obtained for the resting potential seems plausible and is in agreement with qualitative considerations. In the formula the currents of active transport of Na$^+$ and K$^+$ are weighted with the inverse of the respective membrane permeabilities. Because the membrane at rest is much less permeable to Na$^+$ than to K$^+$, the contribution of Na$^+$ is much larger than that of K$^+$.
This weak dependence on the active transport of K$^+$ can explain why the membrane potential changes so little after the poisoning of the ion pumps by an agent like ouabain.

The model calculation also predicts an upper bound for the magnitude of the resting potential, as was found previously on a different mathematical route for an Ohmic-conduction model by Hoppensteadt and Peskin (1992). The upper bound for $|V_m|$ is obtained from a relation between $V_m$ and the intracellular sodium concentration $c_{Na}^{(i)}$ by taking zero as a lower bound for  $c_{Na}^{(i)}$. The relation is in good agreement with data for the giant axon of the squid. The prediction of an upper bound for the absolute magnitude of $V_m$ is consistent with the relatively narrow spread of $V_m$-values observed in different types of cells.

\appendix
\numberwithin{equation}{section}
\section{Comparison with the Ohmic-conduction model of Hoppensteadt and Peskin}

It is of interest to compare with a different model proposed by Hoppensteadt and Peskin (1992). (See also Keener and Sneyd, 1998.) Although the starting expressions look very different,
this model leads, on a different mathematical route, to a similar result of an upper limit to the magnitude of the resting potential. This similarity is gratifying as it suggests a certain robustness of the result for $V_m$ with respect to the choice of the constitutive equations for passive ion flow.

In the model the ion currents of passive transport are described by Ohm's law, as used in the work of Hodgkin and Huxley (1952): 
\begin{equation}
e\Phi_l=g_l(V_m-V_l).
\end{equation}
$e$ is the elementary charge. The conductances $g_l$ are treated as constants. They are the measure of permeability in this model. Only the Nernst potentials $V_l$ depend on the ion concentrations, i.e.
\begin{equation}
V_l=z_l \frac{RT}{F}\ln\left(\frac{c_l^{(o)}}{c_l^{(i)}}\right).
\end{equation}
Hoppensteadt and Peskin (1992) calculated both resting potential and cell volume for this model in the limit $z_i \rightarrow -\infty$.
It leads to the following exponential dependence of the concentration shifts  $(\Delta c_l)_P$ for Na$^+$ and K$^+$ on the pump currents:  
\begin {equation}\label{depletion}
(\Delta c_l)_P=c_l^{(o)}\exp(-F V_m/RT) \left(\exp\left[-\frac{eF}{RT}\frac{\Phi_l^{(a)}}{g_l}\right] -1\right).
\end{equation}
As before in eq.(\ref{DeltaP}), the concentration shift is negative for sodium and positive for potassium, because Na$^+$-ions are pumped out of, and K$^+$-ions are pumped into the cell. ($\Phi_{Na}^{(a)}>0, \Phi_{K}^{(a)}<0$.) 

For low pump rates the dependence of $(\Delta c_l)_P$ is approximately linear, and results are very similar to those obtained with Goldman's electrodiffusion model, with conductances $g_l$ replacing permeabilities $P_l$. For high pump rates, however, the exponential dependence leads to marked differences. As for sodium, the intracellular ion concentration does not go to zero for any finite pump rate. From eq.(\ref{depletion}), the sodium depletion ratio is found as
\begin{equation}
\frac{c_{Na}^{(i)}}{c_{Na}^{(o)}} = exp(-F V_m/RT) exp\left[-\frac{eF}{RT}\frac{\Phi_{Na}^{(a)}}{g_{Na}}\right].
\end{equation}
For potassium, on the other hand, the intracellular ion concentration asymptotically increases without limit for high pump rates. At a certain critical pump rate the condition of stability
\begin{equation}
(\Delta c_+)_P <0
\end{equation}
is violated, which means that the resting potential $V_m$ there goes to zero. Therefore, as a function of the pump rate, $V_
m$ must have a minimum at a certain value $\Phi_{min}$ of $\Phi_{Na}^{(a)}$, somewhere between zero and the limit of stability. The absolute value of $|V_m|$ at the minimum is the upper limit of the magnitude of the resting potential for this model.

I wish to thank Prof. Hans-J\"urgen Apell for many helpful discussions over a long time span.

\section*{References}

\noindent Adam, G., L\"auger, P., Stark G., 2009. Physikalische Chemie und Biophysik. Fifth ed. Springer, Berlin.
,

\noindent Aickin, C.C., Betz, W.J., and  Harris, G.L., 1989. Intracellular chloride and the mechanism for its accumulation in rat lumbrical muscle. J. Physiol. 411, 437-455.



\noindent Cooper, K., Jakobsson, E., Wolynes, P., 1985. The theory of ion transport through membrane channels. Prog.Biophys.molec.Biol. 46, 51-96.

\noindent Gainer, H., Gallant, P.E., Gould, R., Pant, H.C., 1984. Biochemistry and metabolism of the squid giant axon. In: Baker, P.F., ed., Current Topics in Membranes and Transport, vol.22, The squid axon, 57-90. Academic Press, New York.

\noindent Goldman, D.E., 1943. Potential, impedance, and rectification in membranes. J.Gen.Physiol. 27,37-60.

\noindent Gorman, A.L.F., Marmor, M.F., 1970. Contributions of the sodium pump and ionic gradients to the membrane potential of a molluscan neurone. J. Physiol. 210, 897-917.

\noindent Hille, B., 2001. Ion channels of excitable membranes. Third ed. Sinauer. Sunderland.

\noindent Hodgkin, A.L., Huxley, A.F., 1952. The components of membrane conductance in the giant axon of {\it Loligo}. J.Physiol.116,473-496.

\noindent Hodgkin, A.L., Katz, B., 1949. The effect of sodium ions on the electrical activity of the giant axon of the squid. J.Physiol. 108, 37-77.

\noindent Hoppensteadt, F.C., Peskin, C.S., 1992. Mathematics in medicine and the life sciences. Springer, Berlin.

\noindent J\"ackle, J., 2007. The causal theory of the resting potential of cells. J.Theor.Biol. 249, 445-463.


\noindent Katz, B., 1966. Nerve, Muscle, and Synapse. McGraw-Hill, New York.

\noindent Keener, J., Sneyd, J., 1998. Mathematical Physiology. Springer, New York.

\noindent L\"auger, P., 1991. Electrogenic Ion Pumps. Sinauer, Sunderland.

\noindent Mackey, M.C., 1975. Ion Transport through Biological Membranes. Lecture Notes in Biomathematics, vol. 7. Springer, Berlin. (Section 8B).

\noindent Roux, B., 1999. Theories of ion permeation. A chaser. J.Gen.Physiol. 114, 605-608.

\noindent Somjen, G.G., 2004. Ions in the brain: normal function, seizures, and strokes. Oxford University Press, Oxford, New York.


\noindent Stein, W.D., 1990. Channels, carriers, and pumps. An introduction to membrane transport. Academic Press, San Diego CA.

\noindent Tuckwell,  H.C., 1988. Introduction to theoretical neurobiology: vol.1. Cambridge University Press, Cambridge, UK (Chapter 2).

\noindent Woodbury, J.W., 1965. The cell membrane: Ionic and potential gradients and active transport. In: Ruch, T.C., Patton, H.O. (Eds.), Physiology and Biophysics. Saunders, Philadelphia (Chapter 1).

\end{document}